\documentclass[preprint]{aastex}

\newcommand{\be}{\begin{equation}}
\newcommand{\ee}{\end{equation}}

\begin{document}

\title{A Kinship between the EGRET SNR's and Sgr A East} 

\medskip
\author{ Marco Fatuzzo$^1$ and Fulvio Melia$^2$}
\bigskip 
\affil{$^1$Physics Department, Xavier University, Cincinnati, OH 45207} 
\affil{$^2$Physics Department and Steward Observatory, 
The University of Arizona, AZ 85721}

\begin{abstract} 
Sgr A East appears to be a single, mixed-morphology $10,000$-year-old
supernova remnant (SNR) at the Galactic center. It also appears to belong 
to a class of remnants that have been observed and detected at 1720 MHz, 
the transition frequency of OH maser emission. However, if the EGRET 
source 3EG J1746-2852 coincident with the Galactic center is itself 
associated with this object, it would endow it with a $\gamma$-ray 
luminosity almost two orders of magnitude greater than that of the 
other EGRET-detected SNR's. We here reconsider the viability of a
pion-production mechanism as the source of the broadband emission
observed from Sgr A East, and show that what connects these objects---and
ultimately also accounts for their different $\gamma$-ray emissivity---is
the very important interaction between the expanding SNR shell and
the surrounding molecular cloud environment. The singularly high
$\gamma$-ray luminosity of Sgr A East, as well as its unusually
steep radio spectral index, can thereby be attributed to the
high-density ($n_H=10^3$ cm$^{-3}$), strong magnetized
($B\sim 0.18$ mG) environment in which it is located.
\end{abstract}


\section{Introduction}          \label{sec:intro}

Sgr A East is a nonthermal radio source with a supernova
like morphology located near the Galacic center.
Its elliptical structure is elongated
along the Galactic plane with a major axis of length 10.5 pc and a
center displaced from the apparent dynamical nucleus, Sgr A*, by
2.5 pc in projection toward negative Galactic latitudes.  The actual
distance between Sgr A* and the geometric center of Sgr A East
has been estimated to be $\sim 7$ pc  (Yusef-Zadeh \& 
Morris 1987; Pedlar et al. 1989; Yusef-Zadeh et al. 1999).

Recent broadband observations suggest that Sgr A East belongs to
a class of SNR's that have been observed and detected at 1720 MHz, the
transition frequency of OH maser emission (Claussen et al. 1997).  
These observations are consistent with the presence of shocks
produced at the interface between the supersonic outflow 
and the dense molecular cloud environments with which these
SNR's are known to be interacting.
In similar fashion, several maser spots with velocities
of $\approx 50$ km s$^{-1}$ have been resolved in the
region where Sgr A East is interacting with a dense molecular
cloud (the SE boundary), and an additional spot has been 
observed near the Northern arm of Sgr A West (a $\sim 6$-pc mini-spiral
structure of ionized gas orbiting about the center) at a velocity
of $134$ km s$^{-1}$ (Yusef-Zadeh et al. 1996, 1999).  
The SNR-Sgr A East kinship is further suggested
by the EGRET detection of a $\sim 30$ MeV 
- $10$ GeV continuum source (3EG J1746-2852) within the inner 
$1^o$ of the Galactic center (Mayer-Hasselwander et al. 1998), in that
the SNR's observed at 1720 MHz represent a subset of SNR's clearly
associated with EGRET sources (Esposito et al. 1996; Combi, 
Romero \& Bagnaglia 1998; Combi et al. 2001).

There are two caveats with the association of Sgr A East to the EGRET SNR's.
First, nonthermal radio emission from Sgr A East observed at 6 and 20 cm 
is characterized by a spectral index of $\sim 1$, which points to an
underlying population of nonthermal leptons with a power-law distribution 
of index $p \sim 3$ (Pedlar et al. 1989).  
In contrast, typical SNR's have radio emission characterized 
by spectral indeces $\sim 0.5$ and corresponding lepton indeces $\sim 2$.
Second, the implied $\gamma$-ray luminosity of Sgr A East
($\sim 2 \times 10^{37}$ ergs s$^{-1}$) is roughly two orders 
of magnitude greater than the corresponding values for the EGRET
SNR's.  Interestingly, given the 
power required to carve out the radio synchrotron remnant within the 
surrounding dense molecular cloud, it has been argued that the energetics 
of Sgr A East are quite extreme ($\sim 4 \times 10^{52}$ ergs; Mezger 
et al. 1989) compared with the total energy released in a typical supernova 
explosion. We note, however, that this picture may be at odds with the latest 
Chandra X-ray observations of this region (Maeda et al. 2002), which suggest
instead that Sgr A East is the product of a supernova with a typical energy 
release, $\sim 10^{51}$ ergs.  

Adopting the high-energy interpretation, Melia et al. (1998; hereafter MFYM) 
invoked a pion-production 
mechanism within the Sgr A East shell to account for 
the Galactic center EGRET observations. In their scenario, 
the decay of neutral pions created in collisions between shock 
accelerated protons and ambient protons produce a broad spectrum 
that accounts for the observed emission from this source at energies 
greater than $\sim 100$ MeV.  A crucial aspect of this mechanism is 
the decay of charged pions into muons, and subsequently, into 
``secondary" relativistic electrons and positrons. 
The emission from these leptons due to inverse Compton scattering 
with UV photons then accounts for the observed emission at energies 
below $\sim 100$ MeV.  In addition, leptonic synchrotron emission 
also accounts self-consistently for the unusually steep 
($\alpha \sim 1$) radio emission detected from Sgr A East's periphery. 

The work undertaken by MFYM invoked two key elements 
generally ignored in earlier treatments of pion-decay mechanisms 
(see, e.g., Sturner et al. 1997; Gaisser, Protheroe, \& Stanev 1998; 
Baring et al. 1999).  The first is a correct treatment of the energy-dependent
pion multiplicity and its effect on the spectral shape of the pion-decay 
photons (see also Markoff, Melia, \& Sarcevic 1999).  
The second is the self-consistent use of secondary leptons (by-products 
of charged pion decays) in the determination of the broadband emission 
spectrum. The importance of these secondaries becomes significant with 
a proper treatment of the pion multiplicity, as each relativistic proton 
produces on average 40 - 60 leptons.  
As a result, the ratio of secondary leptons to shock-accelerated 
protons is significantly higher than previously 
estimated with an assumed pion-multiplicity of three, bringing 
into question earlier work on pion-production mechanisms that
ignored the role of these decay products.  

Although the scenario put forth in MFYM succesfully describes
the broadband emission of Sgr A East, it does not appear to be
consistent with the conclusion drawn from the recent observations
with ACIS on board the {\it Chandra} X-ray Observatory.  These
show that Sgr A East is probably a single, mixed-morphology 
supernova remnant with an age of $\sim 10,000$ years
(Maeda et al. 2002).  While the {\it Chandra} observation may thus resolve 
the mystery surrounding the birth of Sgr A East, it creates another 
in terms of the relatively large $\gamma$-ray luminosity discussed 
above. Of course, it is quite plausible that 3EG J1746-2852 is not 
associated with Sgr A East. Indeed, it has recently been shown that 
filaments in the arched magnetic field structure to the North of 
the Galactic center may also be viable candidates for the EGRET emission
(Yusef-Zadeh et al. 2002).

Given this new uncertainty, we reconsider
the viability of the scenario put forth by MFYM in the context of 
standard SNR energetics.  Since the EGRET observations fix the neutral 
pion-decay rate ($\sim L_\gamma / 2 / 1$ GeV $\sim 10^{40}$ s$^{-1}$), 
the energy content of the relativistic protons required to produce the
necessary number of neutral pions must be sensitive to the ambient 
proton number density $n_H$; lower values of $n_H$ require a higher 
density of relativistic (shock-accelerated) protons to compensate.  
For example, MFYM found
that an ambient proton density of $10$ cm$^{-3}$ yielded an energy 
content of relativistic particles $\sim 8\times 10^{50}$ ergs.  In contrast, 
an ambient proton density of $10^3$ cm$^{-3}$ yielded an energy 
content $\sim 10^{49}$ ergs.  The view that Sgr A East results from a 
standard SN clearly favors the higher ambient number density.

As the production rates of charged and neutral pions are similar, 
the EGRET observations (which fix the rate of neutral pion production) 
also fix the rate at which secondary leptons (the byproducts of charged 
pions) are produced. (Note, however, that the number of shock 
accelerated protons required to produce these leptons
depends on the pion-multiplicity.)  Indeed, MFYM found this rate to be $\sim 4 
\times 10^{40}$ s$^{-1}$, consistent with the above estimate of 
the neutral pion decay rate. If these particles reach steady-state 
due primarily to cooling via inverse Compton scattering, as was the
case in MFYM, 
then the lepton number density (and hence the emission due to inverse Compton 
scattering and synchrotron processes) is also fixed.
In contrast, bremsstrahlung emission (by the relativistic
particles) scales directly with $n_H$.  As a result, MFYM concluded that 
$n_H$ had to be less than $\sim 30$ cm$^{-3}$ to keep bremsstrahlung 
emission from swamping the EGRET data.

Taken at face value, the above discussion suggests that the scenario 
adopted in MFYM is not compatible with a standard SNR view of Sgr A East.  
However, a steady-state condition for $n_H = 10$ cm$^{-3}$
does not appear to be valid either.
Specifically, the time required to build up the steady-state 
distribution assumed by MFYM is easily estimated to be 
$\approx 7.5 \times 10^6$ years.
It is clear that the amount of time required to build
up a steady-state distribution of secondary leptons far exceeds any 
reasonable age for Sgr A East.

This result implies that MFYM significantly overestimated the emission 
from inverse Compton scattering in their fits to the broadband spectrum, 
making it 
unlikely that this component can account for the low energy portion of the 
EGRET spectrum. However, it is also clear that the lepton number 
density in Sgr A East was also overestimated by as much 
as $\sim 750$, removing the bremsstrahlung constraint that $n_H < 30$
cm$^{-3}$.  In fact, we will show below that 
bremsstrahlung emission can account 
for the EGRET observations at energies below $\sim 100$ MeV
if $n_H \sim 10^3$ cm$^{-3}$.  This value of $n_H$ is 
consistent with the view that Sgr A East is expanding into an 
ionized gas halo with a density $\sim 10^3$ cm$^{-3}$, and interacting 
with a cloud with a density $\sim 10^5$ cm$^{-3}$ (Maeda et al. 2002).  

This paper presents a thorough investigation of this scenario, and is
organized as follows.  An idealized model of 
Sgr A East is presented in \S 2, and then used in \S 3
to determine what relativistic
proton distribution is required to produce a $\pi^0$ emissivity
that can account for the high-energy part of the EGRET observations.
This distribution is then used in \S 4 to calculate the secondary
lepton injection function, and subsequently, to show that the
emission from these secondary leptons can then fill in the low-energy
part of the EGRET observations as well as account for the VLA observations
of Sgr A East.  A summary and conclusion are presented in \S 5.

\section{An Idealized Model for Sgr A East}
Sgr A East is idealized as a uniform spherical shell
with an inner radius of 4 pc and an outer radius of 
5 pc.  The ambient medium is taken to be 
fully ionized hydrogen with an expected density of
$n_H \sim 10^3 - 10^4$ cm$^{-3}$ (see discussion above).  
The region is assumed to be threated by a fairly strong 
(at least $\sim 0.1$ mG) magnetic field (see, e.g., 
Yusef-Zadeh et al. 1996).

Given its proximity to the Galactic center, 
Sgr A East is known to be bathed by the intense IR and UV radiation fields
associated with the central $\sim 1-2$ parsecs of the galaxy
(Telesco et al. 1998; Davidson et al. 1992; Becklin, Gatley \&
Werner 1982).  For simplicity, we adopt values for the
photon energy densities (IR and UV) corresponding to those 
found at a distance of 7 pc from the Galactic center. 
Although the age of Sgr A East is not known, its value has 
been estimated to be $\sim 10^4$ years by Maeda et al. (2002), 
and is expected to be less than 
$\sim 5$ pc/ $100$ km s$^{-1} \sim 5\times 10^4$ 
years, as implied by the observed shock velocities.
As such, we adopt an age of 30,000 years as a reasonable
value consistent with the available constraints for Sgr A East.

Shock accelerated protons are assumed to be uniformly
injected throughout Sgr A East with a power-law distribution.  
These protons cool
primarily through inelastic $pp$ scatterings on a timescale
$\tau_{pp} = [c n_H \sigma_{pp}]^{-1}$, where $\sigma_{pp}$
is the cross-section for proton-proton scattering.  
With $\sigma_{pp} \approx 30$ mbarns (e.g., Sturner et al.
1997), $\tau_{pp}\approx$ 35,000 years for $n_H = 10^3$ cm$^{-3}$.  
For simiplicity, we only consider cases where 
the injected protons in Sgr A East would not have had sufficient
time to cool (see discussion in next section), 
and are therefore described by a distribution
function of the form $n_p(E) = n_o E^{-\alpha}$ (where $E$ is
the total particle energy).  While $\alpha$ is taken as a free 
parameter, its value is expected to range between 2 and 2.4 
(Jones \& Ellison 1991).

\section{The production of $\pi^0$ decay photons}

The $\pi^0$ emissivity resulting from an isotropic
distribution of shock accelerated protons $n_p(E_p)$ 
interacting with cold (fixed target) ambient hydrogen 
of density $n_H$ is given by the expression
\be
Q_{\pi^0}^{pp} = c \; n_H \int_{E_{th}(E_{\pi^0})} 
dE_p\, n_p(E_p) {d\sigma(E_{\pi^0}, E_p)\over dE_{\pi^0}}\;,
\ee
where $E_{th}(E_{\pi^0})$ is the minimum proton energy
required to produce a pion with total energy $E_{\pi^0}$,
and is determined through kinematical considerations.
The resulting $\gamma$-ray emissivity is then
given by the expression
\be
Q_\gamma(E_\gamma) = 2 \int_{E_{\pi^0}^{min} (E_\gamma)}
dE_{\pi^0} {Q_{\pi^0}^{pp} \over (E_{\pi^0}^2 - 
m_{\pi^0}^2 c^4)^{1/2}} \;,
\ee
where $E_{\pi^0}^{min} (E_\gamma) = E_\gamma + m_{\pi^0}^2 c^4 / (4E_\gamma)$.

For simplicity, we shall use the isobar model of Stecker (1973) 
to determine the differential
$\pi^0$ cross-section for proton-proton collisions when the shock
accelerated protons have an energy of less than 3 GeV (e.g., Dermer 1986a,b;
Moskalenko \& Strong 1998).
The lengthy expressions needed to evaluate the cross-section are 
given in the Appendix.  At energies greater than 7 GeV, 
the differential cross-section 
is approximated by the scaling form of Blasi \& Melia (2003; see also
Blasi \& Colafrancesco 1999):  
\be
{d\sigma (E_p, E_{\pi^0})\over dE_{\pi^0}} = {\sigma_0\over E_{\pi^0}} 
f_{\pi^0} (x)\;,
\ee
where $x = E_{\pi^0} / E_p$, $\sigma_0 = 32$ mbarn, 
and 
\be
f_{\pi^0} (x) = 0.67(1-x)^{3.5} + 0.5e^{-18x}\;.
\ee
This scaling form properly takes into account the high pion multiplicities
which occur at high energies. A linear combination of the low and high-energy 
expressions is used in the intermediate 3 - 7 GeV range.

Fits to the EGRET data
for $\alpha$ = 2.0, 2.2, and 2.4, are shown in Figure 1.  The 
fitting was done through the proper choice of the product
$n_H \cdot n_o$, reflecting the fact that an increase in ambient
number density must be offset by a corresponding decrease in
the number of relativistic particles.  Clearly, the total energy budget 
of the relativistic protons is sensitive to the value of $n_H$.
For example, the kinetic energy in relativistic
protons for the $\alpha = 2.2$ case is $\approx 10^{53} (n_H / 1$
cm$^{-3}$) ergs.  A standard SNR interpretion of Sgr 
A East therefore requires high values of $n_H$, although too high
a value then implies significant proton cooling.
We therefore adopt a value of $n_H = 10^3$ cm$^{-3}$
for the remainder of this paper.

It is clear from the fits shown in Figure 1
that neutral pion decay cannot account for the lowest
energy EGRET data.  However, $pp$ scattering also
produces charged pions, which in turn decay into muons, and subsequently,
electrons and positrons.  We consider the production and emission of 
these leptons in the following section.

\section{Secondary lepton production and emission}
Inelastic $pp$ scatterings produce two charged pions for every neutral 
pion.  These charged pions quickly decay into muons, which in turn decay
into positrons and electrons (hereafter referred to collectively as leptons)
with a resulting emissivity
\begin{eqnarray} \nonumber
q_e(E_e) = n_H \; c \; {m_\pi^2 \over m_\pi^2 - m_\mu^2}
\int^{E_\mu^{max}}_{E_\mu^{min}} dE_\mu 
{dP\over dE_e} \int_{E_\pi^{min}}^{E_\pi^{max}}
{dE_\pi \over \beta_\pi E_\pi} \times \\
\int_{E_{th}(E_\pi)} dE_p \; n_p(E_p) 
{d\sigma (E_\pi, E_p) \over dE_\pi}\;.
\end{eqnarray}
Integration limits in the above expression are determined through kinematic 
considerations, and the lepton distribution from 
a decaying muon is given by the three-body decay probability
\begin{eqnarray} \nonumber
{dP\over dE_e} = {8 pc \over \beta_\mu m_\mu^3 c^6}
\int du {u [u^2\gamma_\mu^2 - m_e^2 c^4]^{1/2}
\over (pc - E_e + u)^2} \left(3 - {4\gamma_\mu u
\over m_\mu c^2}\right) \\
\times \left[1 - {E_e(E_e - u)\over p^2 c^2}\right]\;,
\end{eqnarray}
where $u = (E_e - \beta_\mu pc$ cos $\theta)$, $p$ is the electron
momentum, and $\gamma = (1-\beta^2)^{-1/2}$ is the Lorentz factor for 
the designated particle (Markoff, Melia \& Sarcevic 1999).

As with the neutral pion production, we
use the isobaric model to determine the differential cross-section 
for energies less than $3$ GeV (see the Appendix).
For energies greater than $7$ GeV, the differential cross-section
is given by the same expression used for the neutral pion case
(Equation 3),
but with the scaling function multiplied by a factor of two -- that is -- 
$f_\pi (x) = 2 f_{\pi^0} (x)$. A linear combination is used
in the intermediate regime.

The lepton injection rate function $\dot n(\gamma)$
is shown in Figure 2
for a proton spectral index of $\alpha = 2.2$ and the value
of $n_H \cdot n_o$ used in the fit to the EGRET data shown in Figure 1.  
This injection rate is tied directly to the pion production rate,
and hence, to the gamma-ray emission from neutral pion decay. That is,
once the proton spectral index $\alpha$ is chosen, the
lepton injection rate follows directly from fitting
the neutral pion flux to the EGRET data observed at Earth.

The dominant energy loss mechanisms for relativistic leptons with
Lorentz factor $\gamma = [1-\beta^2]^{-1/2}$ in the
Sgr A East environment are:
(1) Synchrotron losses at a rate 
\begin{equation}
{dE_s\over dt} = -{4\over 3} \sigma_T c
\beta^2\gamma^2 {B^2\over 8 \pi}\;;
\end{equation}
(2) Compton scattering losses at a rate
\begin{equation}
{dE_\gamma\over dt} = -{4\over 3} \sigma_T c
\beta^2\gamma^2 \left[{u_{uv}
\sigma_c \left(2.7 x_{uv}\right)
 \over 1 + x_{uv}} +
{u_{ir}
\sigma_c \left(2.7 x_{ir}\right)
\over 1 + x_{ir}}\right]\;,
\end{equation}
where $x_{uv} = \gamma kT_{uv} /m_e c^2$, $x_{ir} = \gamma
kT_{ir} /m_e c^2$, and
\begin{equation}
\sigma_c (x) = {3\over 4} \left\{{1+x \over x^3}\left[
{2x (1+x)\over 1+2x} - \ln (1+2x)\right]
+{1\over 2x} \ln (1+2x) - {1+3x \over (1+2x)^2}\right\}\;;
\end{equation}
(3) bremsstrahlung emission losses at a rate 
\begin{equation}
{dE_B\over dt} = -{3\alpha\over \pi} \sigma_T m_e c^3  \; n_H  
\;\gamma\;\ln (2\gamma - 1/3)\;,
\end{equation}
where $\alpha$ is the fine structure constant;
and (4) Coulomb losses at a rate of
\begin{equation}
{dE_C\over dt} = -{3\over 2} \sigma_T m_e c^3 n_H \beta^{-1}
\lambda_C\;,
\end{equation}
where the Coulomb logarithm $\lambda_C$ is $\approx 24$.
The corresponding cooling rates as a function of $\gamma$,
defined by the general expression $R = E^{-1} (dE/dt)$, 
are plotted in Figure 3 for $n_H = 10^3$ cm$^{-3}$,
$B = 0.18$ mG, and the UV and IR photon fields defined in \S 2
above.

It is clear from Figure 3 that most of the leptons will have had
sufficient time to cool for the assumed age of $3\times 10^4$
years.  We therefore find the lepton distribution function $n(\gamma)$
presently in Sgr A East by solving the steady-state equation
\be
-{\partial\over\partial E_e} \left[n(\gamma) m_e c^2 \left({dE_e\over dt}
\right)_{total}
\right] = q_e (E_e)\;,
\ee
where $n(\gamma) d\gamma = n(E_e) dE_e$.
The result is shown as a dotted curve in Figure 4 (the
significance of the solid line will be discussed below).  

The resulting bremsstrahlung emission from this distribution
of leptons is then given by the corresponding emissivity
\be
Q_b = c\; n_H \int d\gamma \; n(\gamma) \beta {d\sigma\over dE}\;,
\ee
where the lengthy expression for the
differential cross section is taken from Koch \& 
Motz (1959; see also Sturner et al. 1997).  The
resulting bremsstrahlung spectrum observed at Earth is shown as a
dotted curve in Figure 5, with the corresponding spectrum from 
neutral pion decays shown by the short-dashed line.
(The significance of the long-dashed and solid lines will be discussed
below).  

The bremsstrahlung emission at $\sim 100$ MeV is produced primarily by leptons
with $\gamma < 10^3$.  For the physical environment considered
here, these leptons cool primarily via interactions
with the ambient medium (see Figure 3), 
and as such, their steady-state distribution $n(\gamma)$ 
scales as $n_H^{-1}$.  
Under steady-state conditions, the bremsstrahlung 
emissivity is therefore specified solely by $q_e$ (see Equation 5),
and hence is directly tied to the EGRET data.

It is clear that the pion decay mechanism presented here can 
naturally account for the EGRET data in a very robust way.
In stark contrast, however, the resulting population of relativistic
leptons (shown by the dotted line in Figure 4) cannot possibly account 
for the radio observations of Sgr A East.  Specifically, 
the observed radio emission has a spectral index of $\alpha \approx 1$
which in turn requires a power-law lepton distribution with
spectral index $p \approx 3$.  For the assumed field
strength of $B = 0.18$ mG, the leptons which
contribute most to the observed $\lambda$ = 6 cm and 20 cm radio emission
have Lorentz factors of 
\be
\gamma\approx\left({4\pi m_e c^2 \over eB\lambda}\right)^{1/2}
\approx 2 - 4 \times 10^3\;.
\ee
For this range of $\gamma$, the lepton distribution
function is well approximated by a power-law of spectral
index $p \sim 2$, thereby producing synchrotron emission 
with a spectral index of $\sim 0.5$.

One could naively argue that the magnetic fields in the Sgr A East
region are much higher than $0.18$ mG, resulting in cooling rates that scale
as $\gamma^2$ over most of the lepton energies.  Since the injection
rate (see Figure 2) can be characterized as a power law with
spectral index $p \sim 2.2$ above $\gamma\approx 10^2$, it is clear
that the lepton population would then have the desired spectral index of
$p \sim 3$.
However, it is relatively easy to show analytically (using the power-law
expression for synchrotron emission) that the resulting radio flux
in Sgr A East would then be about two orders of magnitude too high.

As such, we make the reasonable assumption that high-energy particles
are diffusing out of the Sgr A East region, and modify the particle
distribution function by a diffusion loss factor $\chi = 
\gamma_0^2 / (\gamma+\gamma_0)^2$.  The parameter $\gamma_0$ reflects
a threshold limit below which diffusion does not significantly
effect the particle population.
While this expression is empirical in nature, it is motivated by the 
fact that the particle gyration radius scales as $\gamma$.  For 
a random walk diffusion, one would then expect that the diffusion
of particles scales as $\gamma^{-2}$ above some threshold value.
(We note that if the diffusion mechanism is due to 
collisions with the ambient medium, it would not be
efficient in normal supernova remnants, as the lower ambient densities
in those regions would dramatically reduce the number of random steps that
leptons could take over the remnant lifetime.)

The diffusion-modified particle distribution function for $n_H
= 10^3$ cm$^{-3}$, $\alpha = 2.2$, $B = 0.18$ mG, and
$\gamma_0 = 5.2 \times 10^3$ is shown by the solid line in 
Figure 4.  Since most of the bremsstrahlung emission at around
$100$ MeV is produced by particles with $\gamma < 10^3$,
the resulting $\gamma$-ray spectrum (represented by the long-dashed
line in Figure 5) is only slightly reduced.  Neverthelss, this
slight reduction yields a better fit to the EGRET data, as
shown by the solid line in Figure 5, which represents the sum
of the neutral pion decay photon spectrum and the diffusion modified
bremsstrahlung spectrum.

The radio emission is easily calculted through the emissivity
\be
Q_s = \left({3^{1/2} e^3\over hmc^2}\right) {B\over h\nu}
\int dE \; n_e(E_e) \; \int_0^{\pi/2} d\theta\; \hbox{\rm sin}^2 \theta
\; F\left({\nu\over \nu_c}\right)\;,
\ee
where 
\be
F(x) = x \int_x^\infty dz K_{5/3} (z)\;,
\ee
and $\nu_c = (3e\gamma^2 B$ sin$\theta / 4\pi m_e c)$.  The
synchrotron emission for the assumed parameters is shown
in Figure 6 along with the 6 cm and 20 cm VLA data
(Pedlar et al. 1989).  The solid line represents the
observed radio flux at Earth from the diffusion-modified particle distribution 
(solid line in Figure 4) and the
dashed line represents the emission from the unmodified 
distribution (dotted line in Figure 4).

\section{Summary and Conclusions}
The recent {\it Chandra} observations of Sgr A East indicate
that this nonthermal Galactic center source is a mixed-morphology
supernova remnant with an age of $\sim 10^4$ years.  In light
of the {\it Chandra} results, we consider the viability of a 
pion-production mechanism as the source of the broadband emission 
observed from Sgr A East.  Our results indicate that such a mechanism
can naturally account for both the EGRET and VLA observations
associated with this source.  The success of this mechanism in
accounting for both radio and $\gamma$-ray emission can be traced to
the proper treatment of the pion multiplicity (as developed
by Markoff, Melia and Sarcevic 1999) which enhances the importance
of secondary leptons with respect to shock accelerated 
protons.  
As these leptons accounted for 
part of the $\gamma$-ray emission observed by EGRET as well as the
radio synchrotron emission observed by the VLA, there is no
need for an assumed distribution of shock accelerated primary
electrons, which we have here assumed to represent a negligible
population relative to that of the secondary particles.  

These results suggest that Sgr A East belongs to a subset of SNR's
that have been observed at 1720 MHz (the transition frequency
of OH maser emission) and that have been associated with EGRET
sources.  What connects these objects physically is their 
interactions with dense molecular cloud environments (Gaisser, 
Protheroe \& Stanev 1998).  The singularly high $\gamma$-ray
luminosity of Sgr A East, as well as its unusually steep radio
spectral index, are then attributed to the high-density ($n_H 
= 10^3$), strong magnetic field ($B = 0.18$ mG) environment in which it
is located.  Although Sgr A East is also bathed by intense 
IR and UV fields, our results indicate that they play a minor role.
That is, Compton cooling of the relativistic leptons is negligible, and
in turn, the emissivity from inverse Compton scattering is considerably
lower than both the bremsstrahlung and neutral pion decay photon
emissivities.  

As a consistency check, we compare the X-ray luminosity 
in the 2 - 10 keV band resulting from bremsstrahlung emission to 
the (model dependent) value obstained by {\it Chandra}
observations.  The former was found to be 
$\sim 10^{32}$ ergs s$^{-1}$, while the latter
is $\sim 8 \times 10^{34}$ ergs s$^{-1}$ for an assumed isothermal
plasma model (Maeda et al. 2002).  It appears then that 
most of the X-ray emission from Sgr A East is emitted by the
thermal ambient plasma rather than the shock accelerated particle
byproducts.

\centerline{\bf Appendix}
We adopt the isobaric model of Stecker (1973) as discussed in 
Dermer (1986a,b) and Moskalenko \& Strong (1999)
to calculate the pion-production
cross-sections for energies less than 3 GeV. In this 
treatment, the differential cross-sections are expressed
as 
\be
{d\sigma(E_p, E_\pi) \over dE_\pi} = \left<\sigma_i (E_p)
\right> {dN(E_p, E_\pi) \over dE_\pi}\;,
\ee
where $ \left<\sigma_i (E_p)\right> $ is the inclusive
cross-section.  The subscript $i$ denotes one of three decay channels:
(1) $pp \rightarrow \pi^0 X$, (2) $pp \rightarrow \pi^+ X$, 
(3) $pp \rightarrow \pi^- X$, where $X$ refers to
the inclusive production of deacy products.
The decay channel $pp \rightarrow \pi^-d$ is not considered here,
as its inclusion would not significantly affect our results.
The inclusive cross-sections for each of these decay
channels are
given by the following parametric expressions (Dermer 1986a):
\begin{eqnarray} \nonumber
 \sigma_{\pi^0 X} = & 0.032 \eta^2
+ 0.040 \eta^6 + 0.047 \eta^8\;,& \hfill p_{p,thr}^{\pi^0 X}
\le p_p \le 0.96\;, \\ \nonumber
&32.6(p_p - 0.8)^{3.21}\;,&\hfill 0.96 \le p_p \le 1.27\;, \\
\nonumber
&5.40(p_p - 0.8)^{0.81}\;,&\hfill 1.27 \le p_p \le 8.0\;, \\
&32.0\; \hbox{\rm ln}(p_p) + 48.5 p_p^{-1/2} - 59.5\;,&
\hfill 8.0 \le p_p 
\;,
\end{eqnarray}
\begin{eqnarray} \nonumber
 \sigma_{\pi^+ X} = & 0.95 \eta^4
+ 0.099 \eta^6 + 0.204 \eta^8\;,& \hfill p_{p,thr}^{\pi^+ X}
\le p_p \le
0.95\;, \\ \nonumber
&0.67 \eta^{4.7} + 0.3\;,&\hfill 0.95 \le p_p \le 1.29\;, \\
\nonumber
&22.0(p_p - 1.27)^{0.15}\;,&\hfill 1.29 \le p_p \le 4.0\;, \\
&27.0\; \hbox{\rm ln}(p_p) + 57.9 p_p^{-1/2} - 40.9\;,&
\hfill 4.0 \le p_p 
\;,
\end{eqnarray}
and
\begin{eqnarray} \nonumber
\sigma_{\pi^- X} = & 
2.33(p_p - 1.65)^{1.2}\;,&\hfill 1.65 \le p_p \le 2.81\;, \\
\nonumber
&0.32 p_p^{2.1}\;,&\hfill 2.81 \le p_p \le 5.52\;, \\
&28.2\; \hbox{\rm ln}(p_p) + 74.2 p_p^{-1/2} - 69.3\;,&
\hfill 5.52 \le p_p 
\;,
\end{eqnarray}
where $p_p$ is the incident (shock-accelerated)
proton momentum in units of GeV/c, $p_{p,thr}^{\pi^0 X} = 0.78$ GeV/c,
and $p_{p,thr}^{\pi^+ X} = 0.80$ GeV/c.

The quantity $\eta$ is given 
by the expression
\be
\eta = {\left[\left(\bar s - m_\pi^2 -  m_\chi^2\right)^2 
-4 m_\pi^2 m_\chi^2
\right]^{1/2}\over 2 m_\pi \bar s^{1/2}}\;,
\ee
where $\bar s = s/c^2 = 2 m_p (E_p/c^2 + m_p)$, and $m_\chi$
depends on the decay channel as follows:  $m_\chi = 
2 m_p$ for the $\pi^0$ decay channel, $m_\chi = 2m_p + m_\pi^+$
for the $\pi^-$ decay channel, and $m_\chi = m_p + m_n$ for the
$\pi^+$ decay channel (where $m_p$ and $m_n$ are the proton and
neutron mass, respectively).
In Equation [21] (as well as those
below), $m_\pi$ is taken as the mass of either neutral or charged
pions, as warranted by the choice of decay channel.

The normalized production spectrum of secondary pions is given
by
\be
{dN(E_\pi, E_p)\over dE_\pi} = w_r(E_p)
\int^{\bar s^{1/2} - m_p}_{m_p+m_\pi} dm_\Delta
B(m_\Delta) f(E_\pi; E_p, m_\Delta)\;,
\ee
where the isobar mass spectrum is given 
by 
\be
B(m_\Delta) = {\Gamma\over\pi}
\left[\left(m_\Delta - m_\Delta^0\right)^2+\Gamma^2\right]^{-1}\;,
\ee
and
\be
w_r(E_p) = \pi \left[\hbox{tan}^{-1}\left({\bar s^{1/2} - m_p - m_\Delta^0
\over \Gamma}\right) - \hbox{tan}^{-1}\left({m_p+m_\pi-m_\Delta^0 \over
\Gamma}\right) \right]^{-1}\;.
\ee
For the $\Delta_{3/2}$(1232), which was the only one considered by Dermer 
(1986b) and, hence, the only one considered here, the resonance width $\Gamma = 
1/2 \times 0.115$ GeV. (The mass and widths of the deltas that decay to
two pions are comparable to those of $\Delta_{3/2}$(1232).) 

The normalized energy spectrum of the pions in the laboratory frame is 
given by
\begin{eqnarray}\nonumber
f(E_\pi; E_p, m_\Delta) =& {1\over 2 m_\pi}
\hbox{\rm \huge\{}\left(2\beta_\Delta^+ 
\gamma_\Delta^+ \beta_\pi'\gamma_\pi'\right)^{-1} 
H\left[\gamma_\pi;\gamma_\Delta^+ \gamma_\pi'(1-\beta_\Delta^+ 
\beta_\pi')\;,
\gamma_\Delta^+ \gamma_\pi'(1+\beta_\Delta^+ \beta_\pi')\right] \\
&+\left(2\beta_\Delta^- 
\gamma_\Delta^- \beta_\pi' \gamma_\pi'\right)^{-1} 
H\left[\gamma_\pi;\gamma_\Delta^- \gamma_\pi'(1-\beta_\Delta^- 
\beta_\pi')\;,
\gamma_\Delta^- \gamma_\pi'(1+\beta_\Delta^- \beta_\pi')\right]
\hbox{\rm \huge\}}
\;,
\end{eqnarray}
where $H[x;a,b] = 1$ if $a \le x \le b$ and $0$ otherwise.  The lab frame
Lorentz factors of the forward (+) and backward (-) isobars are
\be
\gamma_\Delta^\pm  = \gamma_c \gamma_\Delta^* \left(1\pm \beta_c 
\beta_\Delta^*\right)\;,
\ee
where $\gamma_c = \bar s^{1/2} / 2 m_p$ is the center of mass frame
Lorentz factor (with respect to the lab frame), $\gamma_\Delta^*
= (\bar s +m_\Delta^2 - m_p^2)/ (2 \bar s^{1/2} m_\Delta)$ is the
isobar's Lorentz factor in the center of mass frame, and
$\gamma_\pi' = (m_\Delta^2 + m_\pi^2 - m_p^2) / (2 m_\Delta m_\pi)$.

\bigskip 
\centerline{\bf Acknowledgements} 
 
MF is supported by the Hauck Foundation through Xavier University.

\newpage
\begin{figure}
\figurenum{1}
{\epsscale{0.9} \plotone{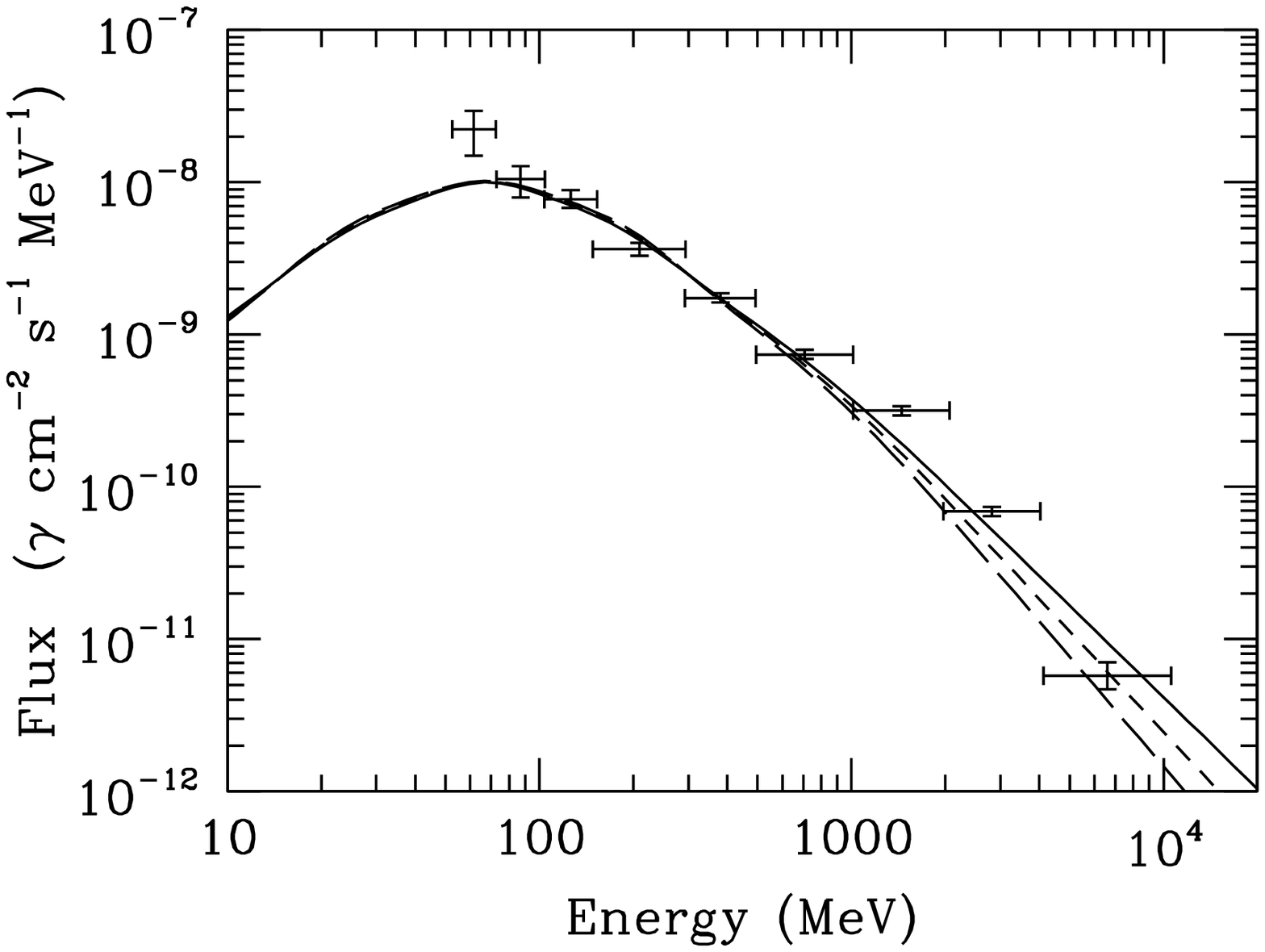} } 
\figcaption{The $\gamma$-ray spectrum due to $\pi^0$ decay
in the Sgr A East shell for three values of the proton
distribution spectral indeces $\alpha$. The solid line corresponds 
to $\alpha = 2.0$, the short-dashed line corresponds to $\alpha
 = 2.2$, and the long-dashed line corresponds to $\alpha = 2.4$.
Each curve was fit to the EGRET data through a proper choice of the
product $n_H \cdot n_o$. The EGRET data are taken from Mayer-Hasselwander
et al. (1999).}
\end{figure}

\newpage
\begin{figure}
\figurenum{2}
{\epsscale {0.9} \plotone{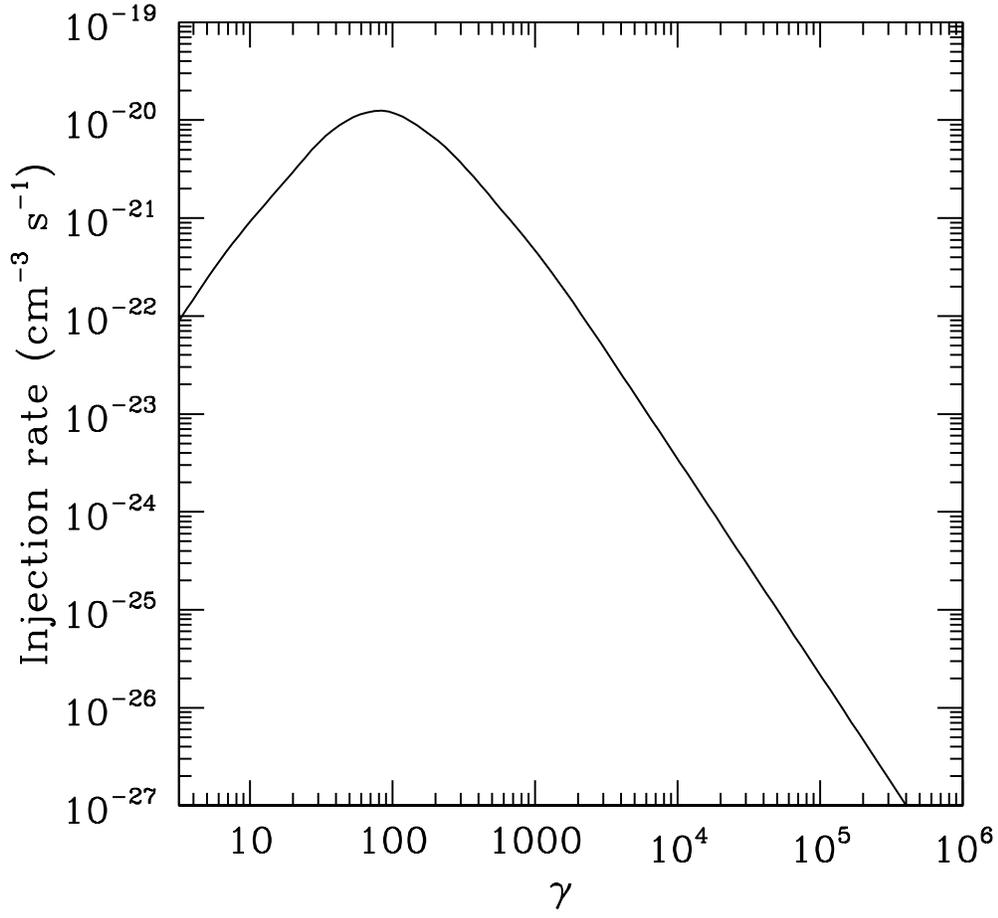} }
\figcaption{The injection rate $\dot n(\gamma)$ for leptons
produced via charged pion decay in Sgr A East for 
$\alpha = 2.2$, assuming the value of $n_H \cdot n_o$ 
used to fit the EGRET data as shown in Figure 1.}
\end{figure}

\newpage
\begin{figure}
\figurenum{3}
{\epsscale{0.9} \plotone{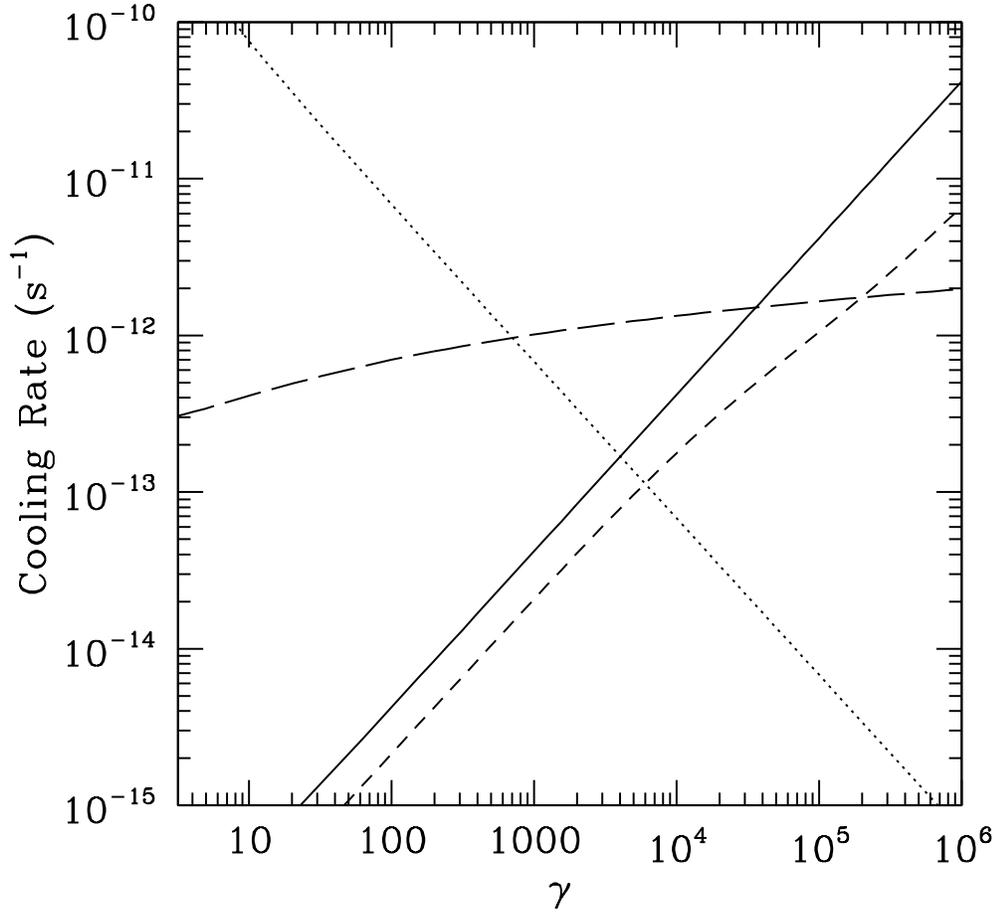} }
\figcaption{Lepton cooling rates  as a function of Lorentz
factor $\gamma$ for $n_H = 10^3$, $B = 1.8$ mG, and the IR and UV
photon fields expected  at a distance of 7 pc from the Galactic
center.  The dotted line represents cooling via Coulomb interactions,
the long-dashed line represents cooling via bremsstrahlung, 
the solid line represents cooling via synchrotron, 
and the short dashed line represents cooling due to Compton scattering.
}
\end{figure}

\newpage
\begin{figure}
\figurenum{4}
{\epsscale{0.9} \plotone{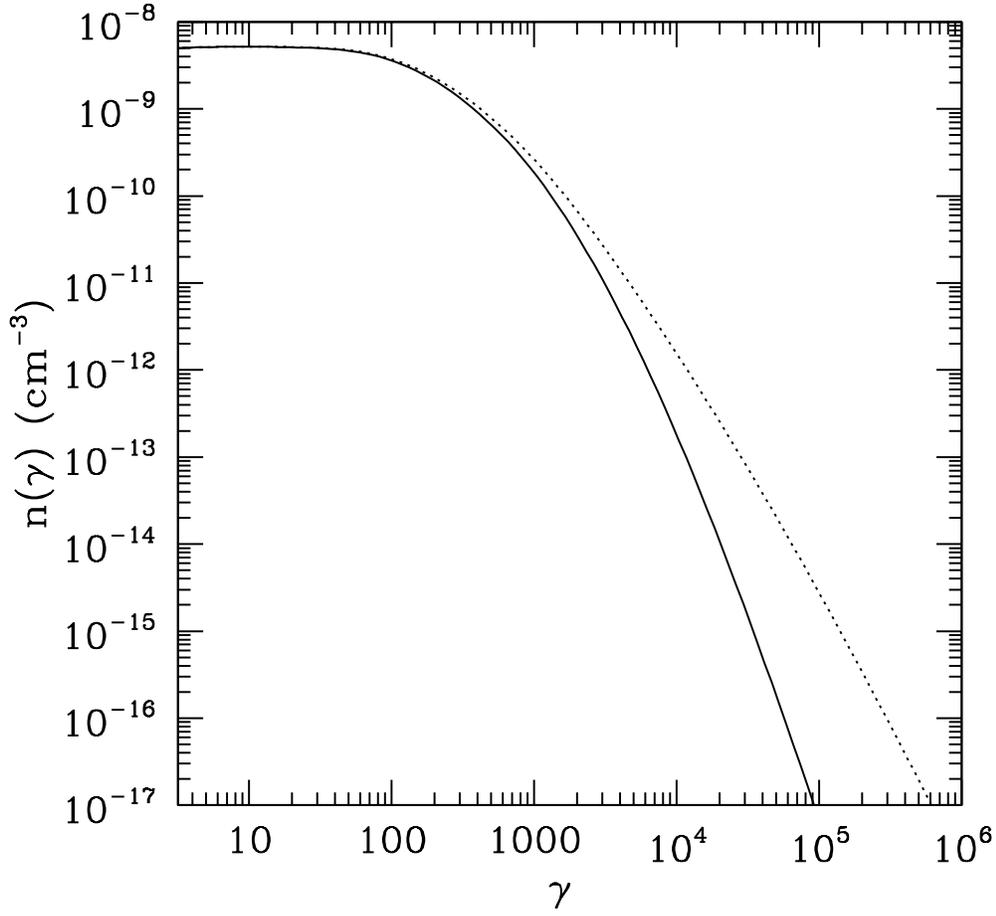} }
\figcaption{The  dotted line represents the
steady-state lepton distribution as a 
function of the particle Lorentz factor $\gamma$ for
the injection rate shown in Figure 2 and the cooling rates 
shown in Figure 3.  The solid line represents the same case, but
with the distribution function modified by a diffusion term
$\chi = \gamma_0^2/(\gamma^2 + \gamma_0^2)$, where $\gamma_0 = 
5.2 \times 10^3$.} 
\end{figure}

\newpage
\begin{figure}
\figurenum{5}
{\epsscale{0.9} \plotone{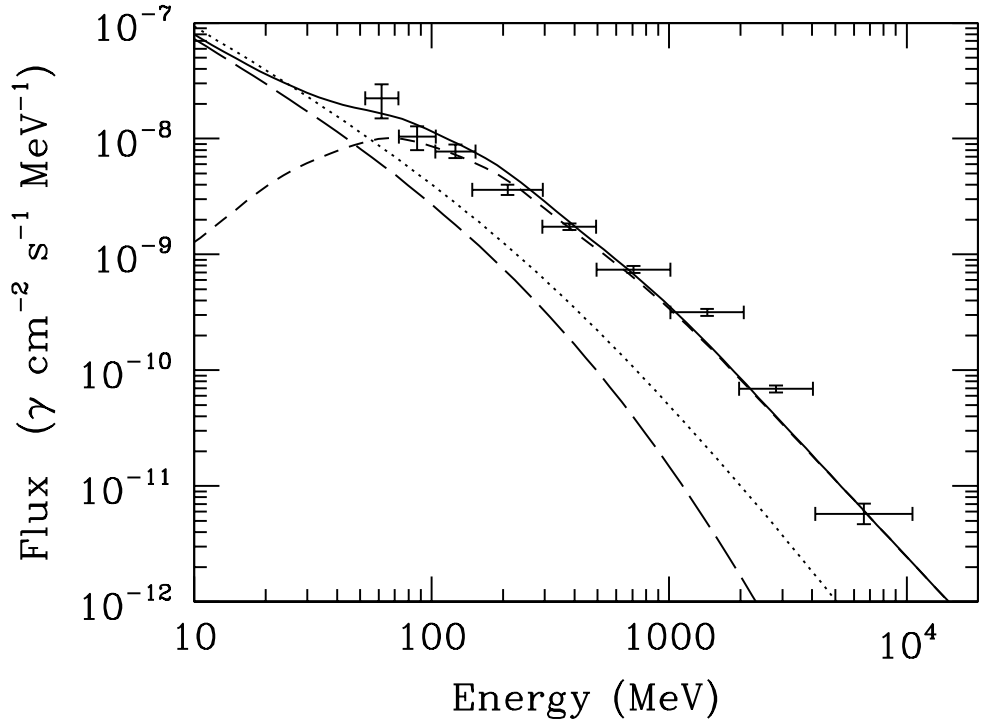} }
\figcaption{The $\gamma$-ray spectrum resulting from bremsstrahlung
and neutral pion decay.  The dotted
line shows the bremsstrahlung emission spectrum resulting from 
the unmodified particle distribution (dotted line
in Figure 4).  The long dashed line shows the same emission spectrum, 
but for the diffusion-modified distribution (solid
line in Figure 4).  The short dashed line shows the $\pi^0$ decay
spectrum for the $\alpha = 2.2$ case shown in Figure 1, and the solid
line is the total spectrum resulting from the bremsstrahlung emission of the
diffusion modified leptons and the $\pi^0$ decays.  The EGRET data are taken 
from Mayer-Hasselwander et al. (1998).
}
\end{figure}

\newpage
\begin{figure}
\figurenum{6}
{\epsscale{0.9} \plotone{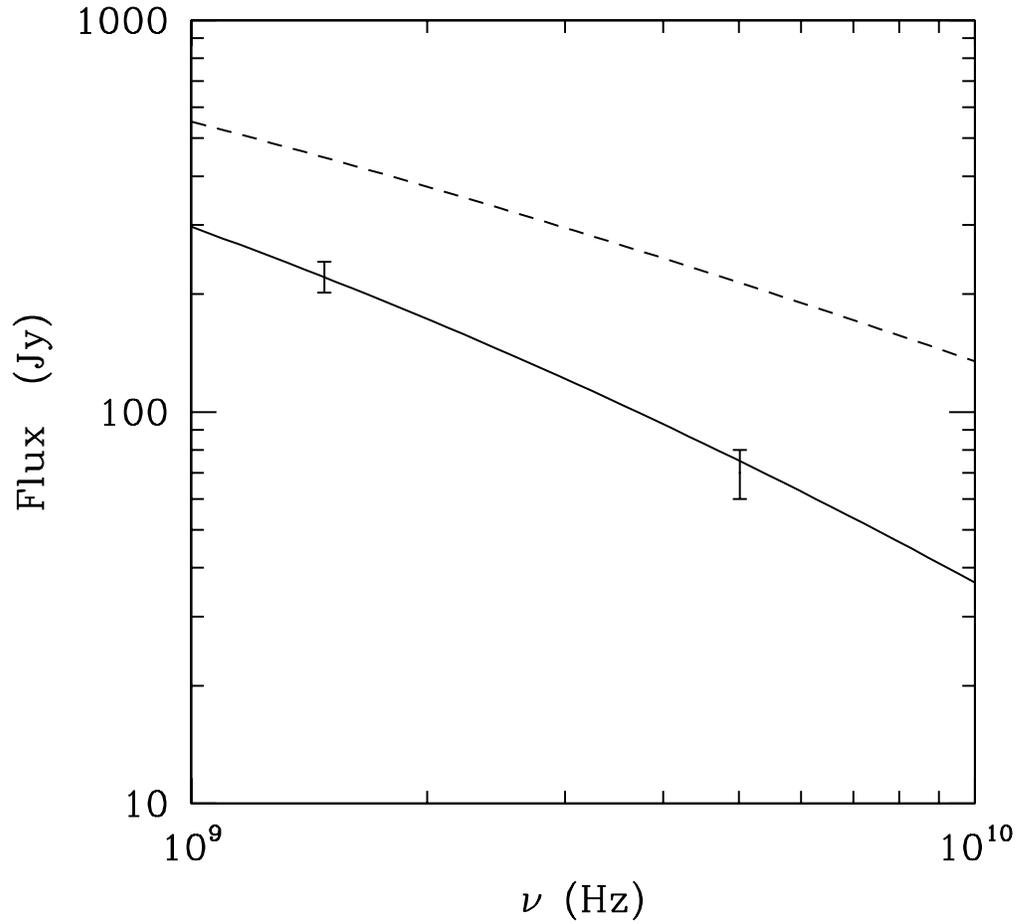} }
\figcaption{The radio flux observed at Earth due to synchrotron emission
of secondary letpons in Sgr A East. The solid line represents the flux
from the diffusion-modified particle distribution shown in Figure 4 
(solid line), and the dashed line shows the flux from the unmodified
distribution shown in Figure 4 (dotted line).  Error bars represent
the VLA observations of the periphery of Sgr A East at 6 cm and 20 cm
(Pedlar et al. 1989).}
\end{figure}
\end{document}